\documentclass[]{aa}
\usepackage{psfig}
\usepackage{rotating}
\begin{document}
\thesaurus{03
           (11.17.1;
            11.17.4 Q0347--383;
            11.17.4 Q0528--250;
            11.17.4 Q0913+072;
            11.17.4 Q1213+093;
            11.08.1)}
\title{On the kinematics of damped Lyman-$\alpha$ systems\thanks{Partly based
on observations carried out at the European Southern Observatory, La Silla,
Chile.}}
\author{C\'edric~Ledoux\inst{1}
   \and Patrick~Petitjean\inst{2,3}
   \and Jacqueline~Bergeron\inst{2,4}
   \and E.Joseph~Wampler\inst{5}
   \and R.~Srianand\inst{6}}
\institute{
   $^1$ Observatoire Astronomique de Strasbourg, 11 Rue de l'Universit\'e, F--67000 Strasbourg, France\\
   $^2$ Institut d'Astrophysique de Paris, 98bis Boulevard Arago, F--75014 Paris, France\\
   $^3$ D.A.E.C, Observatoire de Paris-Meudon, F--92195 Meudon Principal Cedex, France\\
   $^4$ European Southern Observatory, Karl-Schwarzschild Stra$\ss$e 2, D--85748 Garching bei M\"unchen, Germany\\
   $^5$ 418 Walnut Ave., Santa Cruz, CA 95060, USA\\
   $^6$ I.U.C.A.A., Post Bag 4, Ganesh Khind, Pune 411 007, India}
\date{Received 27 April 1998 / Accepted 19 May 1998}
\offprints{C. Ledoux (cedric@astro.u-strasbg .fr).}
\maketitle
\begin{abstract}
We report on high spectral resolution observations of five damped
Lyman-$\alpha$ absorbers: $z_{\rm abs}$ = 3.0248 toward Q 0347--383, $z_{\rm
abs}$ = 2.1411 and 2.8112 toward Q~0528 --250, $z_{\rm abs}$ = 2.6184 toward
Q~0913+072, and $z_{\rm abs}$ = 2.5226 toward Q~1213+093. Line velocity
profiles and heavy element abundances are discussed.

Nitrogen is found to have abundances less than silicon 
in the systems toward Q~0347$-$383, Q~0913+072, and Q~1213+093. The
absorber toward Q~0913+072 is the most metal-deficient damped system known,
with [Fe/H] $<$ $-$3.2. The simple kinematical structure of the metal 
absorptions makes
this system ideal to discuss the [O/Si] and [N/O] ratios. We find 
[O/H] $\approx$~$-$2.7 and $-$2.7~$<$~[Si/H]~$<$ $-$2.2.

By combining these data with information gathered in the literature, we study
the kinematics of the low and high ionization phases in a sample of 26 damped
Lyman-$\alpha$ systems in the redshift range 1.17$-$4.38. We note a strong
correlation between the velocity broadenings of the \ion{Si}{ii}$\lambda$1808
and \ion{Fe}{ii}$\lambda$1608 lines whatever the line optical
depth, implying that the physical conditions are quite homogeneous in the
sample. Statistically this shows that large variations of abundance ratios and
thus large variations of depletion into dust grains are unlikely.
The velocity broadening of the absorption lines, $\Delta V$, is correlated 
with the asymmetry of the lines for $\Delta V$~$<$~150 km~s$^{-1}$. The 
broader the line the more asymmetric it is, as expected in case rotation 
dominates the line broadening. However this correlation does not hold for 
larger $\Delta V$ suggesting that evidence for rotational motions is 
restricted to velocity broadenings $\Delta V$~$<$~150 km~s$^{-1}$. 
The systems with $\Delta V$~$>$~200 km~s$^{-1}$ are peculiar with 
kinematics consistent with random
motions. They show sub-systems as those expected if the objects are in the
process of merging.

Although the sample is not large enough to draw firm conclusions, there is a
trend for the mean velocity broadening to decrease with redshift from
80~km~s$^{-1}$ at $z$~$<$~2.2 to 50~km~s$^{-1}$ at $z$~$>$~2.2.

The kinematics of the low and high ionization species are found to be
statistically correlated, though the high-ionization phase has a much more
disturbed kinematical field than the low-ionization phase. This should be
taken into account in any model of high redshift damped Lyman-$\alpha$
systems.

\end{abstract}

\begin{table*}[tbh]
\caption[]{Journal of observations}
\begin{flushleft}
\begin{tabular}{lccccccc}
\hline
\phantom{Ob}Object&m$_{\rm V}$&$z_{\rm em}$&Date&Instrument&Resolution&Wavelength  &S/N\\
      &           &            &    &          &          &range (\AA )&   \\
\phantom{Obje}(1)&(2)&(3)&(4)&(5)&(6)&(7)&(8)\\
\hline
Q~0347--383& 17.30& 3.230& 20/08/90 & CASPEC & 13000 & 4250--6100 & 12 (25)\\
           &      &      & 16/12/92 & EMMI--R& 33000 & 6100--9900 & 12 (19)\\
Q~0528--250& 17.24& 2.779& 16/01/89 & CASPEC & 13000 & 3650--5100 & 10 (18)\\
           &      &      & 16/12/92 & EMMI--R& 33000 & 5795--9900 & 13 (25)\\
           &      &      & 26/02/95 & CASPEC & 20000 & 3650--5095 & 13 (26)\\
           &      &      & 19/11/95 & CASPEC & 35000 & 3750--4875 & 14 (20)\\
Q~0913+072 & 17.10& 2.785& 16/01/91 & CASPEC & 13000 & 4150--5600 & 15 (37)\\
           &      &      & 16/12/92 & EMMI--R& 33000 & 5800--9900 & 13 (23)\\
           &      &      & 26/02/95 & CASPEC & 20000 & 3900--5095 & 12 (20)\\
Q~1213+093 & 17.20& 2.719& 26/02/95 & CASPEC & 20000 & 3900--5120 & 12 (38)\\
\hline
\end{tabular}
\end{flushleft}
\label{obsetab}
\end{table*}

\section{Introduction}
Damped Lyman-$\alpha$ (hereafter DLA) systems are characterized by a hydrogen
column density $N$(\ion{H}{i})~$>$~10$^{20}$~cm$^{-2}$. The optical depth 
at the Lyman limit is large enough so that hydrogen is neutral. 
The gas is either cold
($T$~$<$~1000 K) or warm ($T$~$\sim$~10$^4$~K) for the highest or lowest
column densities respectively (Petitjean et al. \cite{peti92}). 
As a consequence of the shape of the column density distribution, 
d$^2$$n$/d$N$d$z$~$\propto$~$N^{-\beta}$ with $\beta$~$\sim$~1.5, 
most of the mass is in the systems of highest column densities.
The number density of the latter decreases with time presumably as a
consequence of star-formation (Wolfe et al. \cite{wolf86}; Lanzetta et al.
\cite{lanz95}). Indeed, the cosmic density of neutral hydrogen in DLA
absorbers at $z$~$\sim$~3 is similar to that of stars at the present time
(e.g. Wolfe et al. \cite{wolf95}, Storrie-Lombardi et al. \cite{stor96}; see
also Turnshek \cite{turn97}).

Metallicities and dust content
have been derived from zinc and chromium observations (Meyer et al.
\cite{meye89}; Pettini et al. \cite{pett94}, \cite{pett97a}; Lu et al.
\cite{lu96b}). The [Zn/Cr] ratio
can be considered as an indicator of the presence of dust if it is assumed,
that, as in our Galaxy, zinc traces the gaseous abundances 
whereas chromium is heavily depleted into dust-grains. 
Although these assumptions 
have been questioned by Lu et al. (\cite{lu96b}) and Prochaska \& Wolfe
(\cite{proc97a}), counterarguments have been given by Pettini et al.
(\cite{pett97b}). The typical dust-to-gas ratio determined this way is of the
order of 1/30 of the Milky Way value (Pettini et al. \cite{pett97a};
see also Vladilo 1998). The corresponding
amount of dust could bias the observed number density of DLA systems (Fall \&
Pei \cite{fall93}) and solve the G-dwarf problem (Lanzetta et al.
\cite{lanz95}). Metallicities are of the order of a tenth solar with a
tendency for decreasing metallicity from $z$~$\sim$~2 to $z$~$\ga$~3 (Pettini
et al. \cite{pett97b}; see Boiss\'e et al. \cite{bois98} for lower redshift).
At any redshift however, the scatter is large and it may be hasardous to draw
prematured conclusions from the small sample available. 

Recently, Prochaska \& Wolfe (\cite{proc97b}) have used Keck spectra of 17 DLA
absorbers to investigate the kinematics of the neutral gas using unsaturated
low excitation transitions such as \ion{Si}{ii}$\lambda$1808. They show that
the absorption profiles are inconsistent with models of galactic haloes with
random motions, spherically infalling gas and slowly rotating hot disks. The
CDM model (Kauffmann \cite{kauf96}) is rejected as it produces disks with
rotation velocities too small to account for the large observed velocity 
broadening of the absorption lines.
Models of thick disks ($h$~$\sim$~0.3$R$, where $h$ is the vertical scale and
$R$ the radius) with large rotational velocity ($v$~$\sim$~225~km~s$^{-1}$)
can reproduce the data (see also Prochaska \& Wolfe \cite{proc98}). In a
subsequent paper however, Haehnelt et al. (\cite{haeh98}) use hydrodynamic
simulations in the framework of a standard CDM cosmogony to demonstrate that
the absorption profiles can be reproduced by a mixture of rotational and
random motions in merging protogalactic clumps. The typical virial velocity of
the halos is about 100~km~s$^{-1}$.

The fact that damped systems originate in thick disks 
has been questioned previously. In particular,
the metallicity
distribution of DLA systems is inconsistent with that of stars in the thick
disk of our Galaxy (Pettini et al. \cite{pett97b}; see however 
Wolfe \& Prochaska
\cite{wolf98}). Arguments in favor of DLA systems being associated with 
dwarf galaxies have also been reviewed by Vladilo (1998). However,
it has been shown recently that DLA systems at intermediate redshift are
associated with galaxies of very different morphologies (Le Brun et al.
\cite{lebr97}). 
This strongly suggests that the
objects associated with high-redshift DLA absorbers are progenitors of
present-day galaxies {\sl of all kinds}. 

The discussion of whether high redshift DLA absorption systems are produced in
large, fast rotating disks or in building blocks of galaxies is important
since it is related to how present-day galaxies form, either through initial
formation of large disks and subsequent accretion of gas or as a result of
merging of pregalactic clumps. In this paper, we add observations of five DLA
systems to information gathered in the literature to further address this
problem. We present the observations in Sect.~2, and analyse the velocity
profiles and the abundances in Sect.~3 and 4. In Section~5, we construct an
homogeneous sample of 26 high-redshift DLA systems, and discuss the kinematics
of the low and high ionization species. We draw our conclusions in Sect.~6.

\section{Observations}

\begin{figure}[htb]
\centerline{\hbox{
\psfig{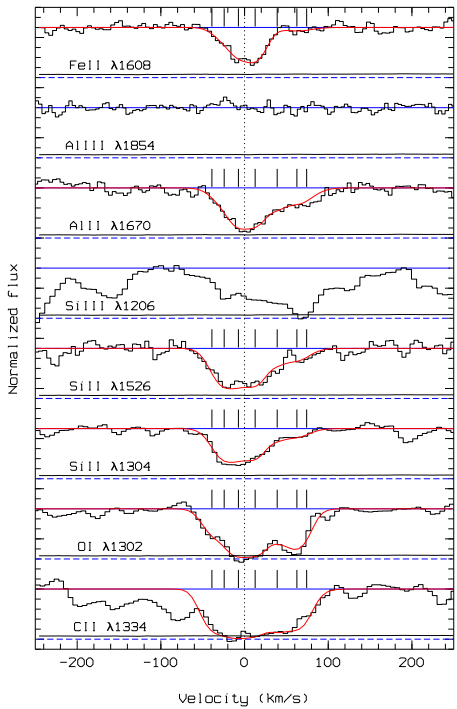}}}
\caption[]{Velocity profiles in the normalized spectra of Q~0347--383: least
square fit of low ionization lines of the $z$~=~3.0248 DLA system. Components
are indicated by a vertical bar. The curve below the spectrum is the noise rms
in the adjacent continuum.}
\label{q0347cin4}
\end{figure}
\begin{figure}[htb]
\centerline{\hbox{\psfig{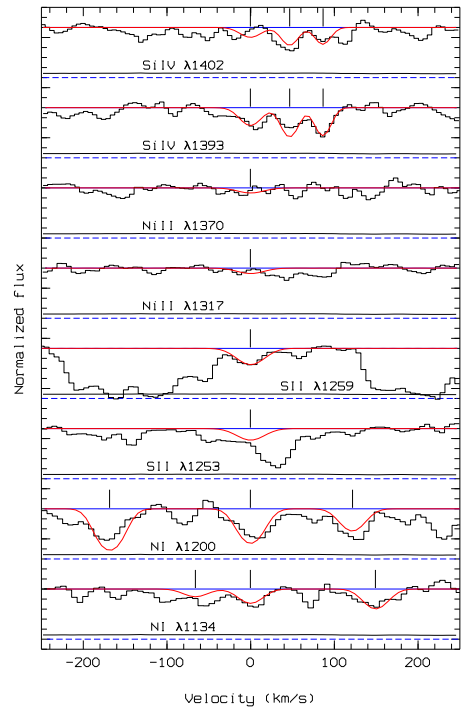}}}
\caption[]{Same as Fig.~\ref{q0347cin4} for the two nitrogen line triplets
and some other low and high ionization species in the $z$~=~3.0248 DLA system
toward Q~0347--383.}
\label{q0347cin5}
\end{figure}

The data were collected at the F/8 Cassegrain focus of the 3.6m telescope, and
at the Nasmyth focus of the 3.5m NTT telescope at the La Silla observatory,
ESO Chile. Blue spectra were obtained 
with the ESO echelle spectrograph (CASPEC). A 300 line mm$^{-1}$ cross 
disperser was used in combination with a 31.6 line mm$^{-1}$ echelle grating. 
For Q~0347--387 and Q~0913+072, the detector was a
RCA CCD with 640x1024 pxl of 15$\mu$m square and a read-out noise of 25
electrons rms. The resolution was $R$~=~13000. Additional data were obtained
on Q~0528--250, Q~0913+072 and Q~1213+093 with higher spectral resolution
($R$~=~20000; $R$~=~35000 for Q~0528--250 as well). In this case, a Tektronix
CCD with 568x512 pxl of 27$\mu$m square and a read-out noise of 10 electrons
was used. Three spectra were obtained through the red arm of the ESO
multi-mode instrument (EMMI) in the echelle spectroscopic mode. A standard
grism was used as cross disperser in combination with the echelle grating \#
10 and a slit width of 2~arcsec, resulting in a resolution of $R$~=~33000. The
detector was a Thomson THX31156 with 1024x1024 pxl of 19$\mu$m square and a
read-out noise of 5 electrons rms. For each exposure on the object, flat-field
images and a wavelength comparison Thorium-Argon spectrum were recorded. The
accuracy in the wavelength calibration measured on the calibrated
Thorium-Argon spectra is about a tenth the resolution quoted above.

The observation log is given in Table~\ref{obsetab}. The magnitudes and quasar
redshifts are taken from Junkkarinen et al. (\cite{junk91}). The mean
signal-to-noise ratio over the wavelength range considered has been computed
from the photon statistics after subtraction of the spectral lines. It is
indicated in the last column together with its maximum value (inside
brackets). The data were reduced using the ECHELLE package implemented within
MIDAS, the image processing system developped at ESO. The cosmic-ray events
have been removed in the regions between object spectra before extraction of
the object. The exposures were co-added to increase the signal-to-noise ratio.
During this merging procedure, the cosmic-ray events affecting the object
pixels were recognized and eliminated. The background sky spectrum was
difficult to extract separately due to the small spacing between the orders in
the blue. Instead, we have carefully fitted the zero level to the bottom of
the numerous saturated lines in the Ly$\alpha$ forest. The uncertainty on the
determination is estimated to be 5\%.

\begin{figure}[bth]
\centerline{\hbox{
\psfig{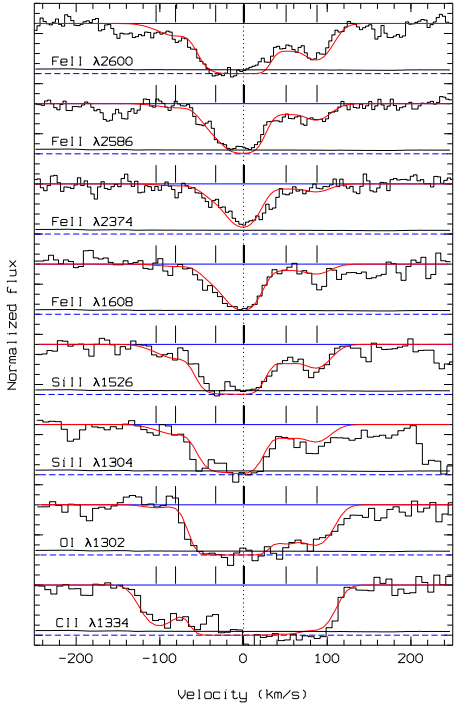}}}
\caption[]{Same as Fig.~\ref{q0347cin4} for low ionization lines in the
$z$~=~2.1411 DLA system toward Q~0528--250.}
\label{q05282cin4}
\end{figure}
\begin{figure}[htb]
\centerline{\hbox{\psfig{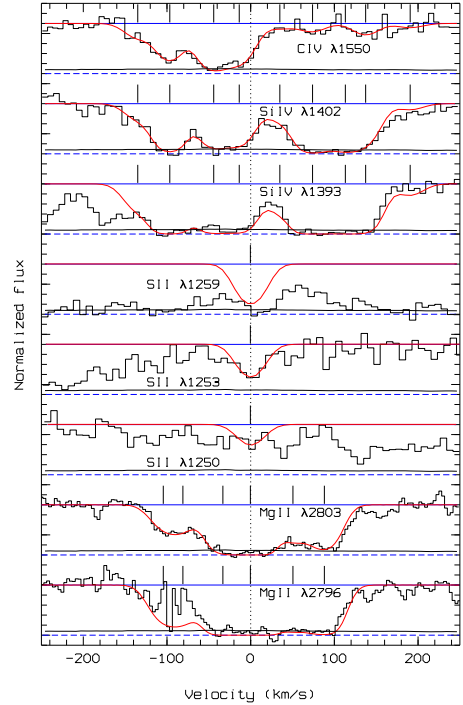}}}
\caption[]{Same as Fig.~\ref{q0347cin4} for the sulfur line triplet and some
other low and high ionization species in the $z$~=~2.1411 DLA system toward
Q~0528--250.}
\label{q05282cin6}
\end{figure}

We have identified all the absorption features with equivalent widths larger
than 5$\times$FWHM$\times$$\sigma$ where $\sigma$ is the noise rms in the
adjacent continuum. Metal line profiles were fitted consistently assuming a
pure turbulent broadening. Ionic column densities were derived with a
least-squares technique using Voigt profiles convolved with the instrumental
spread function, available in MIDAS (FITLYMAN program; Fontana \& Ballester
\cite{font95}). The oscillator strengths are taken from the compilation of
Morton (\cite{mort91}) and the updated values of Savage \& Sembach
(\cite{sava96}). The column densities and the metallicities measured in the
five DLA systems described in the next section are summarized in
Table~2. 
We give the total column densities obtained by summation 
over all the sub-components derived from the fit. 
Metallicities were estimated assuming that neutral and singly ionized species 
were associated with the neutral phase from which the H~{\sc i} 
column density originates.
Upper limits
of detection are computed under the optically thin case approximation. 
When the lines are saturated, we indicate a lower limit obtained by assuming
a mean turbulent broadening
$b$~=~18~km~s$^{-1}$. 
Turbulent $b$ values are given
in the text for lines suspected to lie at the low column density end of the
logarithmic part of the curve of growth; otherwise stated, the line is
optically thin and the resulting column density depends weakly on the
Doppler parameter.
\section{Comments on individual systems}
\subsection{Q~0347--383, $z_{\rm abs}$~=~3.0248}

The damped nature of this Ly$\alpha$ absorber has been first recognized by
Williger et al. (\cite{will89}) with a hydrogen column density
log~$N$(\ion{H}{i})~=~20.8$\pm$0.1~cm$^{-2}$. Pettini et al. (\cite{pett94})
give log~$N$(\ion{H}{i})~=~20.7$\pm$0.1. The strongest part of the absorption
produced by most of the low-ionization species is spread over less than
60~km~s$^{-1}$ consistently centered at $z_{\rm abs}$~=~3.02483 (see
Fig.~\ref{q0347cin4}).
As noted by Junkkarinen et al. (\cite{junk91}), the existence of a strong
\ion{C}{iv} doublet at $z$~=~2.3852 accounting for the
$\lambda$$\lambda$5241.0, 5249.7 features (Steidel \cite{stei90}) is doubtful.
As shown in Fig.~\ref{q0347cin4}, these features can be identified with
\ion{O}{i}$\lambda$1302 and \ion{Si}{ii}$\lambda$1304 respectively, arising
from the DLA system. Moreover neither \ion{C}{ii}$\lambda$1334 nor
\ion{Si}{ii}$\lambda$1260 lines at $z$~$\sim$~2.3852 are detected down to
$w_{\rm obs}$~=~0.14~{\AA } (5$\sigma$). There is also no Ly$\alpha$
associated absorption (Pierre et al. \cite{pier90}; Williger et al.
\cite{will89}).


The \ion{Zn}{ii}$\lambda$2026 and \ion{Cr}{ii}$\lambda$2056 absorptions 
associated with the DLA system have equivalent widths smaller than our 
detection limit, $w_{\rm obs}$~$\la$~0.10 and
0.12~{\AA } respectively at the 5$\sigma$ confidence level (see also Pettini
et al. \cite{pett94}). We marginaly detect \ion{C}{iv}$\lambda$1548 
with $w_{\rm obs}$~=~0.24~{\AA }, while the
\ion{Si}{iv}$\lambda\lambda$1393,1402 doublet is stronger though moderately
weak. \ion{N}{v}$\lambda$1242 (\ion{N}{v} $\lambda$1238 is blended) and
\ion{Al}{iii}$\lambda$1854 are undetected at the 5$\sigma$ upper limits
$w_{\rm obs}$~$=$~0.10 and 0.08~{\AA } respectively. We do confirm the
presence of \ion{N}{i} (Fig.~\ref{q0347cin5}) but with a column density
smaller than previously derived by Vladilo et al. (\cite{vlad97}). By fitting
the two \ion{N}{i} triplets together, we derive log~$N$(\ion{N}{i}) =
14.16$\pm$0.09~cm$^{-2}$ assuming a Doppler parameter $b$ = 19~km~s$^{-1}$.
\subsection{Q~0528--250, $z_{\rm abs}$~=~2.1411}

\begin{figure}[bth]
\centerline{\hbox{
\psfig{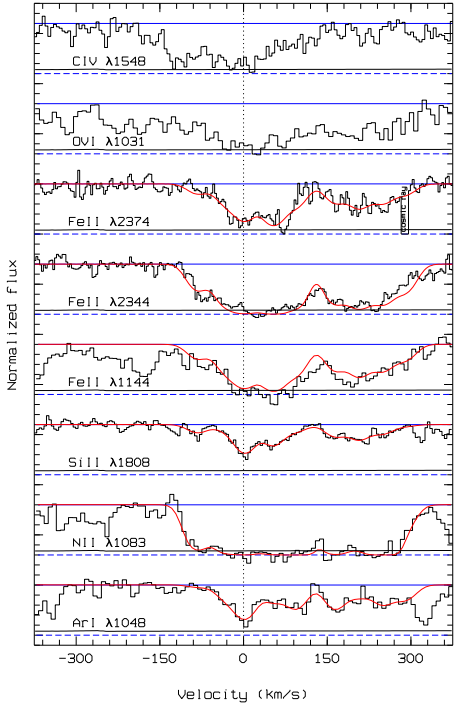}}}
\caption[]{Same as Fig.~\ref{q0347cin4} for metal lines in the $z$~=~2.8112
DLA system toward Q~0528--250.}
\label{q05281cin4}
\end{figure}
\begin{figure}[bth]
\centerline{\hbox{
\psfig{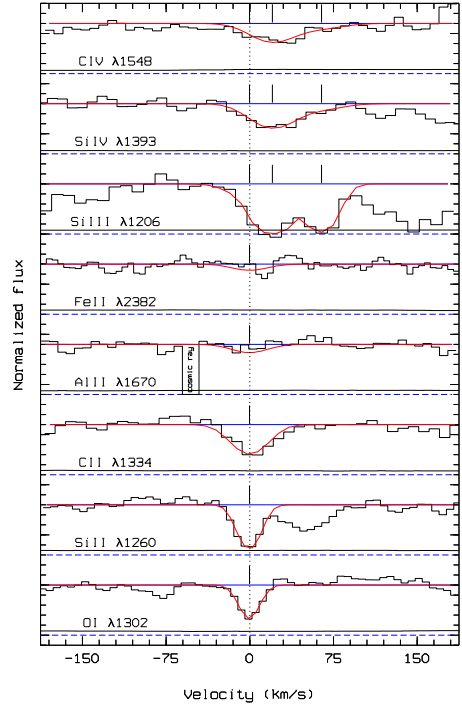}}}
\caption[]{Same as Fig.~\ref{q0347cin4} for low and high ionization species in
the $z$~=~2.6184 DLA system toward Q~0913+072.}
\label{q0913cin4}
\end{figure}

This DLA system has been previously studied by Lu et al. (\cite{lu96b}). We
confirm the hydrogen column density derived by Morton et al. (\cite{mort80}),
log~$N$(\ion{H}{i})~=~20.75 $\pm$0.15~cm$^{-2}$. The \ion{Fe}{ii} column
density is mostly constrained by the \ion{Fe}{ii}$\lambda$2374 line which,
together with the other iron lines available, has been adjusted with a 
Doppler parameter $b$ = 20~km~s$^{-1}$ (see Table~2). The strongest part of
the \ion{Fe}{ii} absorptions has a width of approximately 65~km~s$^{-1}$
(Fig.~\ref{q05282cin4}) whereas the \ion{O}{i}$\lambda$1302 and
\ion{C}{ii}$\lambda$1334 lines are spread over 160~km~s$^{-1}$. The
\ion{Fe}{ii} absorption lie at the red end of the \ion{C}{iv} absorption.
There is a strong \ion{Si}{iv} component centered at +100~km~s$^{-1}$ and
spread over 105~km~s$^{-1}$ that is barely seen in \ion{C}{iv}
(Fig.~\ref{q05282cin6}). For this component only, we derive column densities
in excess of 14.55~cm$^{-2}$ and equal to 13.78$\pm$0.18~cm$^{-2}$ for 
\ion{Si}{iv} and \ion{C}{iv} respectively.
\subsection{Q~0528--250, $z_{\rm abs}$~=~2.8112}

Most of the metal lines have been observed by Lu et al. (\cite{lu96b}). Our
spectrum goes further into the blue however and we show in
Fig.~\ref{q05281cin4} the lines not observed previously at equivalent 
resolution together with a few others for comparison.

This system has drawn much attention because of the high \ion{H}{i} column
density, log~$N$(\ion{H}{i})~=~21.35~cm$^{-2}$ (M\o ller \& Warren
\cite{moll93}) and the detection of H$_2$ molecules, log $N$(H$_2$) =
16.8~cm$^{-2}$ at $z_{\rm abs}$~=~2.8112 (Levshakov \& Varshalovich
\cite{levs85}; Srianand \& Petitjean \cite{sria98}, see also Cowie \& Songaila
\cite{coso95}). The absorption redshift ($z_{\rm abs}$~=~2.8112, as measured 
from the \ion{Ni}{ii} lines by Meyer \& York \cite{meye87} and from the 
H$_2$ lines)
is higher than the QSO emission redshift ($z_{\rm em}$~=~2.77: Morton et al.
\cite{mort80}) by $\sim$3000~km~s$^{-1}$ which makes this system very peculiar
eventhough the QSO emission redshift may have been underestimated. Three
Ly$\alpha$ emission-line objects have been detected within 21~arcsec
from the quasar (or 120~$h^{-1}$~kpc at this redshift and $q_{\rm o}$~=~0.1)
by M\o ller \& Warren (\cite{moll93}) and confirmed by Warren
\& M\o ller (\cite{warr96}) to have redshifts within 200~km~s$^{-1}$ from the
redshift of the DLA system. The width of the
Ly$\alpha$ emission lines is very large ($>$~600~km s$^{-1}$) and continuum
emission could be present (Warren \& M\o ller \cite{warr96}). This suggests
that the gas is not predominantly ionized by the quasar and that
star-formation may occur inside the clouds.

The low \ion{C}{i}/H$_2$ ratio cannot be easily
explained by simple photo-ionization models if solar metallicity ratios are
assumed (Srianand \& Petitjean \cite{sria98}). Indeed,
the low \ion{C}{iv}/\ion{N}{v} ratio and low abundances
($Z/Z_{\odot}$~$\sim$~0.2: Srianand \& Petitjean \cite{sria98}) indicate that
the gas is most certainly not directly associated with the QSO (see Petitjean
et al. \cite{peti94}). The fact that the three Ly$\alpha$ objects are
aligned on the same side of the QSO and have velocities +190, $-$120, +110
km~s$^{-1}$ relative to the damped absorption for projected separations 9, 66
and 115~$h^{-1}$~kpc respectively, argues against the assumption that all the
gas is part of a large rotating disk (Warren \& M\o ller \cite{warr96}). The
system looks like a conglomera of individual clouds. The low and high
ionization absorptions are spread over 440~km~s$^{-1}$ and are organized in
two main sub-systems centered at +40 and +220~km~s$^{-1}$ relative to the
H$_2$ absorption, with a velocity broadening of 220 and 140~km~s$^{-1}$
respectively. It is interesting to note that the high and low ionization
species have similar kinematical structures with well defined sub-components.
Indeed, the weak \ion{N}{v} absorption is at the same redshift as the H$_2$
absorption. This again argues for a group of individual clouds embedded in a
somewhat isotropic ionizing field.
%
\subsection{Q~0913+072, $z_{\rm abs}$~=~2.6184}

\begin{figure}[htb]
\centerline{\hbox{
\psfig{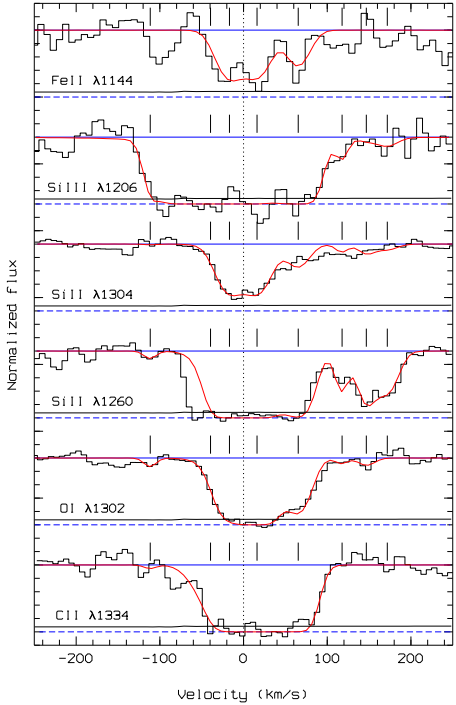}}}
\caption[]{Same as Fig.~\ref{q0347cin4} for low ionization lines in the
$z$~=~2.5226 DLA system toward Q~1213+093.}
\label{q1213cin4}
\end{figure}
\begin{figure}[htb]
\centerline{\hbox{\psfig{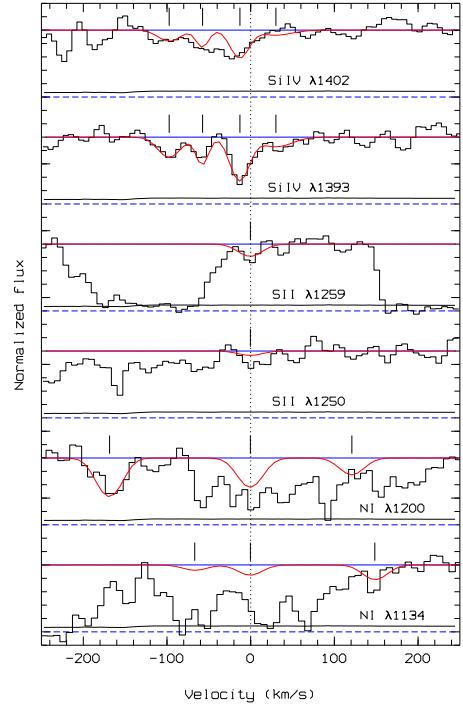}}}
\caption[]{Same as Fig.~\ref{q0347cin4} for the nitrogen and sulfur line
triplets together with the high ionization silicon lines of the $z$~=~2.5226
DLA system toward Q~1213+093.}
\label{q1213cin5}
\end{figure}

From the Ly$\alpha$ line, we derive log~$N$(\ion{H}{i}) = 20.2
$\pm$0.1~cm$^{-2}$ consistent with the measurement by Pettini et al.
(\cite{pett97b}). The low-ionization metal absorptions \ion{O}{i} and
\ion{Si}{ii} show a single and weak component. 
A second component could be present
in the red wing of \ion{C}{ii}$\lambda$1334 (see Fig.~\ref{q0913cin4} in which
the higher resolution data are used for \ion{O}{i} and \ion{Si}{ii}).

The fit of all low-ionization lines together gives a best value
$b$~=~7~km~s$^{-1}$. This means that the lines are barely resolved. The true
value of $b$ could thus be smaller and the column densities larger than those
indicated in Table~2. {\sl However}, the non-detection of
\ion{Si}{ii}$\lambda$1304 puts strong constraints on $b$. Indeed in order for
this line to be unseen in our data, the \ion{Si}{ii} column density {\sl
cannot be larger than} 13.58 whatever the $b$ value is. For smaller $b$
values, the fit would be inconsistent with the stronger \ion{Si}{ii} lines.
The lower limit for the \ion{Si}{ii} column density is 10$^{13.1}$~cm$^{-2}$.
If it is assumed that the above $b$ value is the same for \ion{O}{i} then the
\ion{O}{i} column density is well determined. Although the data are consistent
with a single \ion{O}{i}, \ion{Si}{ii} component, we cannot completely rule
out the possibility that the \ion{Si}{ii} line, in reality, spans a slightly
larger velocity range than \ion{O}{i} which would imply a smaller $b$ value
for \ion{O}{i} than for \ion{Si}{ii}.

It is important to note that the \ion{Fe}{ii}$\lambda$2382 line is
{\sl not} detected down to $w_{\rm obs}$~$<$~0.15~{\AA } at the 5$\sigma$
confidence level.

We do not detect high ionization lines (\ion{Si}{iii}, \ion{Si}{iv} and
\ion{C}{iv}) at the velocities of the low ionization lines. Rather, the high
ionization absorptions lie at +21 and +65 km~s$^{-1}$ relatively to the low 
ion absorptions.
%
\subsection{Q~1213+093, $z_{\rm abs}$~=~2.5226}
%
%
We confirm the damped nature of this system with a hydrogen column density
log~$N$(\ion{H}{i})~=~20.1$\pm$0.1~cm$^{-2}$. The redshift derived from the
\ion{Si}{ii}$\lambda$1304 line is $z_{\rm abs}$ = 2.5226 (the strongest
part of the absorption is spread over less than 60~km~s$^{-1}$,
see Fig.~\ref{q1213cin4}). There is an additional satellite component
detected by \ion{Si}{ii} absorption at +145~km~s$^{-1}$. This satellite 
system is however barely seen in \ion{O}{i}$\lambda$1302 and not seen at all 
in \ion{C}{ii}$\lambda$1334. The nitrogen triplets are heavily blended (see
Fig.~\ref{q1213cin5}) but the \ion{N}{i}$\lambda$1199 line, which leads
to a low column density, log~$N$(\ion{N}{i}) =
13.86$\pm$0.06 cm$^{-2}$, for $b$ = 18~km~s$^{-1}$.
Down to $w_{\rm obs}$~=~0.12~{\AA } (5$\sigma$), we do not detect the
\ion{N}{v}$\lambda$1242 line associated with the possible
\ion{N}{v}$\lambda$1238 absorption mentioned by Sargent et al.
(\cite{sarg88}). 

\section{Metallicity}\label{metallicity}
The metal and dust contents of DLA systems have been estimated in the past few
years using weak lines of zinc and chromium (Meyer et al. \cite{meye89};
Pettini et al. \cite{pett94}). The ratio [Cr/Zn] is assumed to be an 
indicator of the depletion of refractory elements into dust as zinc is not 
depleted in the ISM of our Galaxy whereas chromium is.

\begin{table*}[tbh]
{\scriptsize {\small\caption[]{Metal abundances}}
\begin{flushleft}
\begin{tabular}{lrrrrrrrrrr}
\hline
    & \multicolumn{2}{c}{Q~0347--383 ($z$~=~3.02)} & \multicolumn{2}{c}{Q~0528--250 ($z$~=~2.14)}
    & \multicolumn{2}{c}{Q~0528--250 ($z$~=~2.81)} & \multicolumn{2}{c}{Q~0913+072   ($z$~=~2.61)} & \multicolumn{2}{c}{Q~1213+093 ($z$~=~2.52)}\\
Ion & \multicolumn{1}{c}{log~$N$} & [Z/H]$^{\star}$ & \multicolumn{1}{c}{log~$N$} & [Z/H]$^{\star}$
    & \multicolumn{1}{c}{log~$N$} & [Z/H]$^{\star}$ & \multicolumn{1}{c}{log~$N$} & [Z/H]$^{\star}$ & \multicolumn{1}{c}{log~$N$} & [Z/H]$^{\star}$\\
&\multicolumn{1}{c}{(cm$^{-2}$)}&&\multicolumn{1}{c}{(cm$^{-2}$)}&&\multicolumn{1}{c}{(cm$^{-2}$)}&&\multicolumn{1}{c}{(cm$^{-2}$)}&&\multicolumn{1}{c}{(cm$^{-2}$)}&\\
(1)&\multicolumn{1}{c}{(2)}&(3)\phantom{.6}&\multicolumn{1}{c}{(4)}&(5)\phantom{.6}&\multicolumn{1}{c}{(6)}&(7)\phantom{.6}&\multicolumn{1}{c}{(8)}&(9)\phantom{.6}&\multicolumn{1}{c}{(10)}&(11)\phantom{.6}\\
\hline
\ion{H}{i}          &     20.70$\pm$0.10$^{a}$          & ...... &     20.75$\pm$0.15\phantom{$^{a}$}&  ......  &     21.35$\pm$0.10$^{b}$          & ...... &     20.20$\pm$0.10\phantom{$^{a}$}&    ......&     20.10$\pm$0.10\phantom{$^{a}$}&    ......\\
\ion{C}{ii}         &  $>$14.87\phantom{$\pm$0.00$^{a}$}&$>$-2.38&  $>$15.34\phantom{$\pm$0.00$^{a}$}&  $>$-1.96&  $>$15.58$^{c}$\phantom{$\pm$0.00}&$>$-2.32&  $>$14.08\phantom{$\pm$0.00$^{a}$}&  $>$-2.67&  $>$15.36\phantom{$\pm$0.00$^{a}$}&  $>$-1.29\\
\ion{C}{iv}         &     13.15$\pm$0.13\phantom{$^{a}$}& ...... &  $>$14.75$^{\dagger}$\phantom{$\pm$0.00}&  ......  &$>$15.00$^{c}$\phantom{$\pm$0.00}& ...... &   13.54$\pm$0.11\phantom{$^{a}$}&......&  ........\phantom{$\pm$0.00$^{a}$}&......\\
\ion{N}{i}          &     14.16$\pm$0.09\phantom{$^{a}$}&   -2.51&  ........\phantom{$\pm$0.00$^{a}$}&  ......  &  ........\phantom{$\pm$0.00$^{a}$}& ...... &  $<$13.28\phantom{$\pm$0.00$^{a}$}& $<$-2.89&     13.86$\pm$0.06\phantom{$^{a}$}&     -2.21\\
\ion{N}{ii}         &  $<$14.73\phantom{$\pm$0.00$^{a}$}& ...... &  ........\phantom{$\pm$0.00$^{a}$}&  ......  &  $>$15.60\phantom{$\pm$0.00$^{a}$}& ...... &  $<$13.70\phantom{$\pm$0.00$^{a}$}&    ......&  ........\phantom{$\pm$0.00$^{a}$}&    ......\\
\ion{N}{v}          &  $<$13.35\phantom{$\pm$0.00$^{a}$}& ...... &  ........\phantom{$\pm$0.00$^{a}$}&  ......  &     13.99$\pm$0.04$^{c}$& ...... &  $<$12.98\phantom{$\pm$0.00$^{a}$}&    ......&  $<$13.82\phantom{$\pm$0.00$^{a}$}&    ......\\
\ion{O}{i}          &  $>$15.16\phantom{$\pm$0.00$^{a}$}&$>$-2.41&  $>$15.62\phantom{$\pm$0.00$^{a}$}&  $>$-2.00&  $>$15.94$^{c}$\phantom{$\pm$0.00}&$>$-2.28&     14.28$\pm$0.14\phantom{$^{a}$}&     -2.79&  $>$15.59\phantom{$\pm$0.00$^{a}$}&  $>$-1.38\\
\ion{O}{vi}         &  ........\phantom{$\pm$0.00$^{a}$}& ...... &  ........\phantom{$\pm$0.00$^{a}$}&  ......  &  $<$15.37\phantom{$\pm$0.00$^{a}$}& ...... &  ........\phantom{$\pm$0.00$^{a}$}&    ......&  ........\phantom{$\pm$0.00$^{a}$}&    ......\\
\ion{Mg}{ii}        &  ........\phantom{$\pm$0.00$^{a}$}& ...... &  $>$14.61\phantom{$\pm$0.00$^{a}$}&  $>$-1.72&  $<$15.88$^{c}$\phantom{$\pm$0.00}&$<$-1.05&  ........\phantom{$\pm$0.00$^{a}$}&    ......&  ........\phantom{$\pm$0.00$^{a}$}&    ......\\
\ion{Al}{ii}        &  $>$13.13\phantom{$\pm$0.00$^{a}$}&$>$-2.05&  $>$13.46$^{c}$\phantom{$\pm$0.00}&  $>$-1.77&  $>$14.20$^{c}$\phantom{$\pm$0.00}&$>$-1.63&  $<$11.94\phantom{$\pm$0.00$^{a}$}&  $<$-2.74&  ........\phantom{$\pm$0.00$^{a}$}&    ......\\
\ion{Al}{iii}       &  $<$12.04\phantom{$\pm$0.00$^{a}$}& ...... &     12.77$\pm$0.08$^{c}$          &  ......  &  $>$14.07$^{c}$\phantom{$\pm$0.00}& ...... &  $<$12.27\phantom{$\pm$0.00$^{a}$}&    ......&  ........\phantom{$\pm$0.00$^{a}$}&    ......\\
\ion{Si}{ii}        &  $>$14.41\phantom{$\pm$0.00$^{a}$}&$>$-1.84&     15.26$\pm$0.04$^{c}$          &     -1.04&     16.05$\pm$0.10\phantom{$^{a}$}&   -0.85&     13.34$\pm$0.24\phantom{$^{a}$}&     -2.41&  $>$14.74\phantom{$\pm$0.00$^{a}$}&  $>$-0.91\\
\ion{Si}{iii}       &  $<$13.58\phantom{$\pm$0.00$^{a}$}& ...... &  ........\phantom{$\pm$0.00$^{a}$}&  ......  &  ........\phantom{$\pm$0.00$^{a}$}& ...... &  $<$13.98\phantom{$\pm$0.00$^{a}$}&    ......&  $>$15.88\phantom{$\pm$0.00$^{a}$}&    ......\\
\ion{Si}{iv}        &     13.45$\pm$0.12\phantom{$^{a}$}& ...... &  $>$14.46$^{\dagger}$\phantom{$\pm$0.00}&  ......  &$>$14.52$^{c}$\phantom{$\pm$0.00}& ...... &   13.29$\pm$0.11\phantom{$^{a}$}&......&13.50$\pm$0.16\phantom{$^{a}$}&......\\
\ion{S}{ii}         &     14.46$\pm$0.04\phantom{$^{a}$}&   -1.51&  $<$15.08\phantom{$\pm$0.00$^{a}$}&  $<$-0.94&     15.59$\pm$0.03$^{c}$          &   -1.03&  $<$13.92\phantom{$\pm$0.00$^{a}$}&  $<$-1.55&  $<$14.22\phantom{$\pm$0.00$^{a}$}&  $<$-1.15\\
\ion{Ar}{i}         &  ........\phantom{$\pm$0.00$^{a}$}& ...... &  ........\phantom{$\pm$0.00$^{a}$}&  ......  &    $<$14.53\phantom{$\pm$0.00$^{a}$}& $<$-1.24&  ........\phantom{$\pm$0.00$^{a}$}&    ......&  ........\phantom{$\pm$0.00$^{a}$}&    ......\\
\ion{Cr}{ii}        &  $<$12.86\phantom{$\pm$0.00$^{a}$}&$<$-1.52&     13.10$\pm$0.04$^{c}$          &     -1.33&     13.65$\pm$0.12$^{c}$          &   -1.38&  $<$12.53$^{d}$\phantom{$\pm$0.00}&  $<$-1.35&  ........\phantom{$\pm$0.00$^{a}$}&    ......\\
\ion{Mn}{ii}        &  ........\phantom{$\pm$0.00$^{a}$}& ...... &     12.38$\pm$0.10$^{c}$          &     -1.90&  ........\phantom{$\pm$0.00$^{a}$}& ...... &  $<$12.27\phantom{$\pm$0.00$^{a}$}&  $<$-1.46&  ........\phantom{$\pm$0.00$^{a}$}&    ......\\
\ion{Fe}{ii}        &  $>$14.25\phantom{$\pm$0.00$^{a}$}&$>$-1.96&     14.68$\pm$0.07\phantom{$^{a}$}&     -1.58&     15.45$\pm$0.11$^{c}$          &   -1.41&  $<$12.43\phantom{$\pm$0.00$^{a}$}&  $<$-3.28&  $<$14.61\phantom{$\pm$0.00$^{a}$}&  $<$-1.00\\
\ion{Ni}{ii}        &  $<$12.94\phantom{$\pm$0.00$^{a}$}&$<$-2.01&     13.22$\pm$0.06$^{c}$          &     -1.78&     13.89$\pm$0.03$^{c}$          &   -1.71&  $<$13.02\phantom{$\pm$0.00$^{a}$}&  $<$-1.43&  $<$13.02\phantom{$\pm$0.00$^{a}$}&  $<$-1.33\\
\ion{Zn}{ii}        &  $<$12.15\phantom{$\pm$0.00$^{a}$}&$<$-1.20&  $<$12.28$^{c}$\phantom{$\pm$0.00}&  $<$-1.12&     13.09$\pm$0.07$^{c}$          &   -0.91&  $<$11.90$^{d}$\phantom{$\pm$0.00}&  $<$-0.95&  ........\phantom{$\pm$0.00$^{a}$}&    ......\\
\hline
\end{tabular}
\end{flushleft}
{\small $^{\star}$ Metallicity relative to the solar value from Savage \&
Sembach (\cite{sava96}); $^{\dagger}$ Fit of the blue component only (see text
for explanations); $^{a}$ Pettini et al. \cite{pett94}; $^{b}$ M\o ller \&
Warren \cite{moll93}; $^{c}$ Lu et al. \cite{lu96a}; $^{d}$ Pettini et al.
\cite{pett97b}.}
\label{abuntab}}
\end{table*}

Some controversy has arisen recently about the presence of dust in DLA
systems. Lu et al. (\cite{lu96b}) have argued that nucleosynthesis alone can
explain the element abundance ratios observed in DLA systems and that the
presence of dust is thus questionable. They claim that the relative abundances
are consistent with the bulk of heavy elements being produced by Type~II
supernovae. Pettini et al. (\cite{pett97b}) have shown however that, even
though the nucleosynthesis history of the gas must play a role, the depletion
levels of the refractory elements indicate the presence of dust.
The subsequent level of extinction (Kulkarni et al. 1997; Vladilo 
\cite{vlad98}) is in qualitative
agreement with the direct extinction measurement by Pei et al. (\cite{pei91}).
We discuss here in turn the three most interesting absorbers 
toward Q~0347--383, Q~0913+072 and Q~1213+093.

In the ISM of our Galaxy, sulfur is not heavily depleted into dust grains
(Sembach \& Savage \cite{semb96}). The gas-phase metallicity in the
$z_{\rm abs}$~=~3.0248 system toward Q~0347--383 is thus of the order of
$-$1.5. The silicon and iron abundances are of the same order of magnitude,
[Si/Fe]~$\sim$~0 and [S/Fe]~$\sim$~0.5. This is consistent with solar relative
abundances and little depletion into dust. Nickel seems to be slighly depleted
as compared to iron. It is interesting to note however that, in our Galaxy,
stars with [Fe/H]~$\sim$~$-$1, have [Si/Fe] $\sim$ [S/Fe] $\sim$ 0.4 (e.g.
Fran\c cois \cite{fran87}) and [Ni/Fe] $\sim$~0 (Gratton \& Sneden
\cite{grat91}). This suggests that indeed the nucleosynthesis history of the
damped systems may be different from that of our Galaxy and that the presence
of dust at such a low level may be difficult to disentangle from
nucleosynthesis effects.

The most important observation in this system comes from \ion{N}{i}. The
nitrogen to silicon abundance ratio is smaller than solar
([N/Si] $<$ $-$0.65). 
A similar abundance pattern is observed in the $z_{\rm abs}$~=~2.6184
absorber toward Q~0913+072 (see below), and in the $z_{\rm abs}$ = 2.5226
absorber toward Q~1213+093 (see Table~2).
Such low values of the relative nitrogen abundance have been
observed as well at $z_{\rm abs}$ = 2.2794, 2.309, 2.5379 and 2.8443 toward
Q~2348--147, Q~0100+130, \object{Q~2344+124} and Q~1946+769 respectively
(Pettini et al. \cite{pett95}; Molaro et al. \cite{mola98}; Lipman
\cite{lipm95}; Lu et al. \cite{lu95}, see also Lu et al. \cite{lu98}). It is
important to recall that the ionization correction factor for nitrogen is
always close to unity for log~$N$(H~{\sc i})~$>$~20 cm$^{-2}$ (Viegas
\cite{vieg95}).

The system toward Q~0913+072 is characterized by low metallicities and simple
kinematical structure. This implies that absorptions from usually heavily
saturated lines (\ion{O}{i}$\lambda$1302 and \ion{C}{ii}$\lambda$1334) can be
used to derive abundances. The data are consistent with [C/H] $\approx$ [O/H]
$\approx$ $-$2.7 although the fit of the lines gives a Doppler parameter $b$
$\sim$ 7~km~s$^{-1}$, which indicates that the lines are barely resolved in
our spectrum and may be slighly saturated. We have argued in Sect.~3.4 that
$-$2.7~$<$~[Si/H] $<$~$-$2.2 which implies that [Si/O] might be slighly
over-solar. The iron abundance is surprizingly low, [Fe/H]~$<$~$-$3.2, which,
if confirmed, would be the smallest value amongst known DLA
systems. Moreover, [Fe/Si] $\le$ $-$0.8. The abundance pattern is similar to
that observed in the $z_{\rm abs}$ = 2.076 system toward Q~2206--199
(Prochaska \& Wolfe \cite{proc97a}) in which [Si/H] $\sim$ $-$2.2 and [Fe/Si]
= $-$0.4. In the framework of chemical evolution models, and assuming no
depletion into dust grains, these numbers can be explained if the system is in
the early phase of a burst of star-formation. In such models, however, the
oxygen abundance is expected to be twice that of silicon (Matteucci et al.
\cite{matt97}). It would be interesting to investigate whether such an
abundance pattern is similar to that in Ly$\alpha$ forest clouds with
log~$N$(\ion{H}{i}) $\sim$ 15 cm$^{-2}$ (Cowie et al. \cite{cowi95}). 

The [N/Si] ratio is found to be less than solar. Given the low
absolute metallicity, this system should be studied in more detail to investigate
the nucleosynthesis history of DLA systems in the framework presented by Lu
et al. (\cite{lu98}). Indeed, although the resolution of our data is slighly
too low to definitively conclude, it seems that [O/Si] may be slighly over-solar.

\section{The metal line profiles of DLA systems}
Recently, Prochaska \& Wolfe (\cite{proc97b}) have studied the velocity
profiles of unsaturated low ionization metal transitions (mostly
\ion{Si}{ii}$\lambda$1808) of 17 DLA systems. They conclude that the observed
structures are consistent with a model of rapidly rotating, thick disk and
inconsistent with disk models from a CDM structure formation scenario
(Kauffmann \cite{kauf96}). Haehnelt et al. (\cite{haeh98}) have shown however
that the data can be reconciled with such models if a detailed treatment of
the gas dynamics is included. Here we want to address the problem of the
kinematics without any a priori model in mind. We emphasize that large
velocity broadenings arise most often in peculiar systems and that high and
low-ionization species show some correlation in their kinematics.

We have compiled a sample of 26 damped Ly$\alpha$ systems from the literature
(see Table~\ref{kinetab} for references). For each of them we have measured
the velocity broadening of the profiles $\Delta V$ for the low and high
ionization species at different optical depths. $\Delta V_{\rm low}^{\rm r}$
is the velocity width of a low-excitation line measured as the velocity
separation of the wavelengths at which the residual in the normalized spectrum
is $r$. We consider $r$~=~0.9 and 0.5.
%
Except for a few cases, there is a redshift at which the optical depth of
unsaturated low-excitation transitions (\ion{Zn}{ii}, \ion{Fe}{ii}) reaches a
maximum. We take this redshift as the origin of velocities. $V_{1}$ ($>$~0)
and $V_{2}$ ($<$~0) are the largest and smallest velocities for which the
residual in the normalized spectrum is $r$. We define an asymmetry parameter
$\eta$~=~$\left|V_1+V_2\right|/(V_1-V_2)$ that is close to zero for a
symmetric profile and close to one for a one-sided profile. The results are
presented in Table~\ref{kinetab}. We emphasize that the velocity 
interval $\Delta V$ considered here is larger than that of 
Prochaska \& Wolfe (\cite{proc97b}).
These authors have chosen to ignore 5\% of the integrated optical depth at
each edge of the profiles to correct for the internal broadening of the
components. For high level of asymmetry,
our definition of $\eta$ leads consequently to slightly smaller values of this
parameter. After rescaling however, the distributions of the asymmetry
parameter considered in both paper are similar (see Sect.~\ref{velhisto}). 
In Column~12 of
Table~\ref{kinetab}, we give the measurements of d$_{l\leftrightarrow h}$
which is the velocity difference between the maximum of the optical depth
for the low and high-ionization species respectively.
\begin{table*}[tbh]
\caption[]{DLAS kinematics measurement}
\begin{flushleft}
\begin{tabular}{lccccccccccc}
\hline
\phantom{Ob}Object&$z_{\rm abs}$&log~$N$(\ion{H}{i})&Ref.& Transition & $\Delta$V$^{0.9}_{low}$ & $\Delta$V$^{0.5}_{low}$ & $\eta ^{0.5}_{low}$
& Transition & $\Delta$V$^{0.9}_{high}$ & $\eta ^{0.9}_{high}$ & d$_{l\leftrightarrow h}$\\
   &             &(cm$^{-2}$)       &    &            & (km\,s$^{-1}$)          & (km\,s$^{-1}$)          &
&            & (km\,s$^{-1}$)           &                      & (km\,s$^{-1}$)          \\
\phantom{Obje}(1)&(2)&(3)&(4)&(5)&(6)&(7)&(8)&(9)&(10)&(11)&(12)\\
\hline
\object{Q~0450--132} & 1.17& ... &  1 & \ion{Fe}{ii}$\lambda$2344 & 205& 140& 0.71&                      ... & ...&  ...& ...\\
\object{Q~0449--134} & 1.27& ... &  1 & \ion{Fe}{ii}$\lambda$2586 &  85&  55& 0.27&                      ... & ...&  ...& ...\\
\object{Q~0935+417}  & 1.37&20.30&  2 & \ion{Fe}{ii}$\lambda$2586 & 102&  47& 0.49&                      ... & ...&  ...& ...\\
\object{Q~1946+769}  & 1.74& ... &  1 & \ion{Fe}{ii}$\lambda$2344 &  52&  34& 0.29& \ion{C}{iv}$\lambda$1548 &  97& 1.10&  25\\
\object{Q~0216+080}  & 1.77&20.00&  1 & \ion{Fe}{ii}$\lambda$2586 &  95&  40& 0.00&                      ... & ...&  ...& ...\\
\object{Q~1331+170}  & 1.78&21.18&  3 & \ion{Fe}{ii}$\lambda$1608 &  95&  75& 0.60& \ion{C}{iv}$\lambda$1548 & 375& 0.01& 130\\
\object{Q~2206--199} & 1.92&20.65&  4 & \ion{Fe}{ii}$\lambda$1608 & 203& 153& 0.70& \ion{C}{iv}$\lambda$1548 & 300& 0.50&  32\\
\object{Q~0201+365}  & 1.95&20.18&  5 & \ion{Fe}{ii}$\lambda$1608 & 245& 205& 0.17&                      ... & ...&  ...& ...\\
\object{Q~2231--0015}& 2.07&20.56&  1 & \ion{Fe}{ii}$\lambda$1608 & 170& 113& 0.77&                      ... & ...&  ...& ...\\
\object{Q~2206--199} & 2.08&20.43&  4 & \ion{Fe}{ii}$\lambda$1608 &  15& ...&  ...& \ion{C}{iv}$\lambda$1548 &  62& 0.19&   7\\
\object{Q~0528--250} & 2.14&20.70&  1 & \ion{Fe}{ii}$\lambda$1608 & 160&  65& 0.38& \ion{C}{iv}$\lambda$1548 & 187& 0.44&  43\\
\object{Q~2348--147} & 2.28&20.57&  6 & \ion{Si}{ii}$\lambda$1304 &  40&  30& 0.00&                      ... & ...&  ...& ...\\
\object{Q~0216+080}  & 2.29&20.45&  1 & \ion{Fe}{ii}$\lambda$1608 & 160&  95& 0.58& \ion{C}{iv}$\lambda$1548 & 500& 0.48&   5\\
\object{Q~0100+130}  & 2.31&21.40&  7 & \ion{Fe}{ii}$\lambda$1608 &  77&  55& 0.64& \ion{C}{iv}$\lambda$1548 & 103& 0.26&  50\\
\object{Q~0201+365}  & 2.46&20.38&  5 & \ion{Fe}{ii}$\lambda$1608 & 235& 215& 0.02& \ion{C}{iv}$\lambda$1550 & 360& 0.78&  80\\
\object{Q~1213+093}  & 2.52&20.10&  8 & \ion{Si}{ii}$\lambda$1304 & 217&  72& 0.03&\ion{Si}{iv}$\lambda$1393 & 172& 0.49&  20\\
\object{Q~0913+072}  & 2.62&20.20&  8 & \ion{Si}{ii}$\lambda$1526 &  20& ...&  ...&\ion{Si}{iv}$\lambda$1393 &  97& 0.59&  25\\
\object{Q~0528--250} & 2.81&21.20&  1 & \ion{Fe}{ii}$\lambda$1608 & 400& 325& 0.63& \ion{C}{iv}$\lambda$1550 & 445& 0.21&  25\\
\object{Q~1425+606}  & 2.83&20.30&  1 & \ion{Fe}{ii}$\lambda$1608 & 170&  30& 0.00& \ion{C}{iv}$\lambda$1548 & 445& 0.46& 225\\
\object{Q~1946+769}  & 2.84&20.27&  1 & \ion{Fe}{ii}$\lambda$1608 &  14& ...&  ...& \ion{C}{iv}$\lambda$1548 & 205& 0.02&  72\\
\object{Q~0347--383} & 3.02&20.50&  8 & \ion{Fe}{ii}$\lambda$1608 & 115&  43& 0.16&\ion{Si}{iv}$\lambda$1393 & 125& 0.52&  40\\
\object{Q~2233+131}  & 3.15&20.00&  9 & \ion{Fe}{ii}$\lambda$1608 & 235& ...&  ...& \ion{C}{iv}$\lambda$1548 & 300& 0.53& 150\\
\object{Q~0000--263} & 3.39&21.41&  1 & \ion{Fe}{ii}$\lambda$1608 &  65&  30& 0.00& \ion{C}{iv}$\lambda$1548 & 280& 0.14&  50\\
\object{Q~2212--1626}& 3.66&20.20&  1 & \ion{Fe}{ii}$\lambda$1608 &  80& ...&  ...& \ion{C}{iv}$\lambda$1548 &  68& 0.32&  17\\
\object{Q~2237--0608}& 4.08&20.48&  1 & \ion{Fe}{ii}$\lambda$1608 & 160& ...&  ...& \ion{C}{iv}$\lambda$1548 & 202& 0.14&  35\\
\object{Q~1202--0725}& 4.38&20.60& 10 & \ion{Fe}{ii}$\lambda$1608 & 110& ...&  ...& \ion{C}{iv}$\lambda$1548 &  43& 1.09&  25\\
\hline
\end{tabular}
\end{flushleft}
REFERENCES:
(1) Lu et al. \cite{lu96b}; (2) Meyer et al. \cite{meye95}; (3) Wolfe
\cite{wolfe}; (4) Prochaska \& Wolfe \cite{proc97a}; (5) Prochaska \& Wolfe
\cite{proc96}; (6) Pettini et al. \cite{pett95}; (7) Wolfe et al.
\cite{wolf94}; (8) This work; (9) Lu et al. \cite{lu97}; (10) Lu et al.
\cite{lu96a}.
\label{kinetab}
\end{table*}
\subsection{The \ion{Fe}{ii}$\lambda\lambda$1608,2586 and 
\ion{Si}{ii}$\lambda$1808 profiles}
The velocity broadening of the \ion{Fe}{ii}$\lambda\lambda$1608,2586 (at
$r$~=~0.5 or $\tau_{\nu}$~$\sim$~0.7) and \ion{Si}{ii}$\lambda$1808 (at
$r$~=~0.9 or $\tau_{\nu}$~$\sim$~0.1) line profiles are compared in
Fig.~\ref{patsfss}. It can be seen that they correlate very closely. This
implies that we can use the iron velocity interval that is easier to measure
because of larger optical depths. The iron lines are observed in a large
number of systems and we are able to compare velocities measured {\sl with the
same ion}. Prochaska \& Wolfe (\cite{proc97b}) made used of several
transitions because \ion{Si}{ii}$\lambda$1808 line is not detected in all the
systems.

Since the correlation is very good, it implies that the physical conditions
are quite homogeneous in the sample. First we can be confident that the
criteria are well choosen to probe the densest regions where iron and silicon
are both predominantly singly ionized. If it were not the case we would expect
large variations in the $N$(\ion{Si}{ii})/$N$(\ion{Fe}{ii}) ratio as
\ion{Si}{ii} persists much longer than \ion{Fe}{ii} when the \ion{H}{i} column
density decreases. Indeed the $N$(\ion{Si}{ii})/$N$(\ion{Fe}{ii}) ratio can be
larger than ten for log~$N$(\ion{H}{i})~$\sim$~18~cm$^{-2}$ (see Petitjean et
al. \cite{peti92}). It also suggests that large variations of abundance ratios
and/or large depletion into dust grains are unlikely. These conclusions are
consistent with detailed studies of DLA systems (Lu et al. \cite{lu96b},
Prochaska \& Wolfe \cite{proc96}, Pettini et al. \cite{pett97b}, Vladilo
1998). We can infer
from this correlation that, statistically, the local optical depth of
\ion{Fe}{ii}$\lambda\lambda$1608,2586 is approximately seven times larger than
the local optical depth of \ion{Si}{ii}$\lambda$1808. Hence, using the
oscillator strengths given by Savage \& Sembach (\cite{sava96}), the
\ion{Si}{ii} column density is on average three times larger than the
\ion{Fe}{ii} column density. The solar metallicities of iron and silicon are
similar which implies that in these systems and on average,
[Si/Fe]~$\sim$~0.5 in the gas phase. 

\begin{figure}[ht]
\centerline{\hbox{
\psfig{figure=0870.f9,height=7.0cm,rheight=6.5cm}}}
\caption[]{The velocity broadening of the
\ion{Fe}{ii}$\lambda\lambda$1608,2586 lines, measured at an optical depth
$\tau_{\nu}$~$\sim$~0.7, versus that of the \ion{Si}{ii}$\lambda$1808 line
(measured at $\tau_{\nu}$~$\sim$~0.1) for the whole sample.}
\label{patsfss}
\end{figure}

%
\subsection{The \ion{Fe}{ii} velocity broadening distribution}\label{velhisto}
\begin{figure*}[bth]
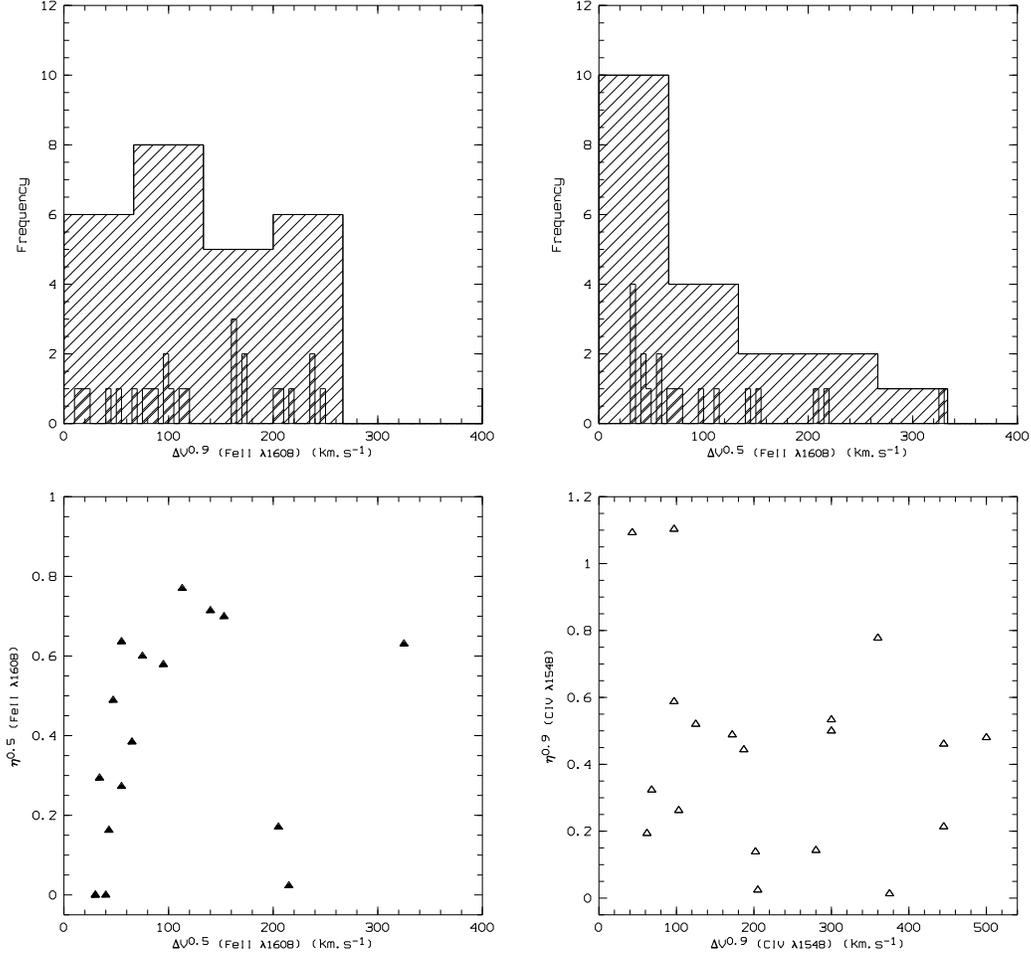

\centerline{\hbox{
\psfig{figure=0870.f10,height=7.0cm,rheight=6.5cm}
\psfig{figure=0870.f11,height=7.0cm,rheight=6.5cm}}}
\centerline{\hbox{
\psfig{figure=0870.f12,height=7.0cm,rheight=6.8cm}
\psfig{figure=0870.f13,height=7.0cm,rheight=6.8cm}}}
\caption[]{{\sl Upper panels:} histograms of the velocity broadening of the
\ion{Fe}{ii}$\lambda\lambda$1608,2586 lines measured at an optical depth 
$\tau_{\nu}$~$\sim$~0.1 ({\it
upper left}) and $\tau_{\nu}$~$\sim$~0.7 ({\it upper right}) for the whole
sample of 26 DLA systems; the smaller binning indicates 
individual systems.
{\sl Lower panels:} asymmetry parameter (see text) versus velocity 
broadening for
the \ion{Fe}{ii} lines measured at $\tau_{\nu}$~$\sim$~0.7 ({\it lower left})
and for the high-ionization lines measured at $\tau_{\nu}$~$\sim$~0.1 ({\it
lower right}).}
\label{pats_f}
\end{figure*}
Fig.~\ref{pats_f} (upper panels) shows the histograms of the velocity
broadening measured from \ion{Fe}{ii} at $\tau _{\nu}$~$\sim$~0.1 
(left panel) and $\tau_{\nu}$~$\sim$~0.7 (right panel). When weak wings 
are considered ($\tau _{\nu}$~$\sim$~0.1), the distribution is
nearly uniform from 20 to 250~km~s$^{-1}$. This is what is found by Prochaska
\& Wolfe (\cite{proc97b}). However, when only the deepest part of the
absorption is considered (and we have given arguments showing that this part
should be associated with the neutral phase), the distribution is peaked at
small velocity broadenings. Most of the profiles have
$\Delta v$~$<$~100~km~s$^{-1}$, while a few have
$\Delta v$ $>$ 150 km~s$^{-1}$. It can be seen in Fig.~\ref{pats_f} (lower
left panel) that the asymmetry parameter $\eta$ is correlated with the
velocity broadening up to $\Delta V$~$\sim$~150~km~s$^{-1}$. For larger
velocity broadenings however this correlation disappears. This approach
complements the edge-leading test of Prochaska \& Wolfe (\cite{proc97b};
Fig.~10) as it takes into account the correlation between 
asymmetry and velocity broadening that is expected in 
fast rotating thick-disk models.

Since it is the presence of systems with large velocity
broadenings that mostly rules out the CDM model (see Fig.~10 of Prochaska \&
Wolfe \cite{proc97b}), we discuss below these systems individually:\\

%
$\bullet$ {\it Q~0201+365 at $z$~=~2.462:} If we strictly follow our
procedure and measure the velocity broadening of 
\ion{Fe}{ii} $\lambda$1608 at $\tau$~=~0.5, we find $\Delta
V$~$\sim$~600~km~s$^{-1}$. This large velocity spread is due to a satellite of
the main system, which has a column density 
log~$N$(\ion{H}{i})~$\sim$~19~cm$^{-2}$ (Prochaska \& Wolfe \cite{proc96}). 
In this case, since the velocity separation is quite large, there is 
little doubt that we should consider that this satellite
arises in a system that is {\sl not} associated with the main damped component.
However this case clearly shows that
the presence of satellites (as expected in the case of merging
processes) could bias the discussion on the amount of rotation needed to
explain the profiles.
%
By restricting our consideration to the main system, the velocity broadening
is $\Delta V$~$\sim$~225~km~s$^{-1}$. The system does not show the
edge-leading pattern. Actually the Ly$\alpha$ line is fitted with 23
components (Prochaska \& Wolfe \cite{proc96}), none of which has
log~$N$(\ion{H}{i})~$>$~20~cm$^{-2}$ and four of which have
log~$N$(\ion{H}{i})~$>$~19~cm$^{-2}$. This system is definitively atypical.\\

$\bullet$ {\it Q~0201+365 at $z$~=~1.955:} The \ion{Fe}{ii}$\lambda$1608
criterion gives $\Delta V$~$\sim$~230~km~s$^{-1}$. Here again there is no
obvious edge-leading pattern. Even the Ly$\alpha$ line is poorly fitted by a
damped component and the system resembles much more the LLS system at
$z$~=~2.325, where strong \ion{Fe}{ii} components are spread over
200~km~s$^{-1}$, whereas the Ly$\alpha$ line does not show damped wings (see
Prochaska \& Wolfe \cite{proc96}).\\

$\bullet$ {\it Q~0528--250 at $z$~=~2.811:} The system is at
$z_{\rm abs}$~$>$~$z_{\rm em}$ but is certainly not associated with the quasar
(Warren \& M\o ller \cite{warr96}; Srianand \& Petitjean \cite{sria98}). It
has $\Delta V_{FeII}^{0.5}$~= 325~km~s$^{-1}$ and the profile can be
decomposed in two main sub-systems, each with no edge-leading
structure. We follow Prochaska \& Wolfe (\cite{proc97b}) and do not consider
it as a typical damped system.\\

\begin{figure*}[bth]
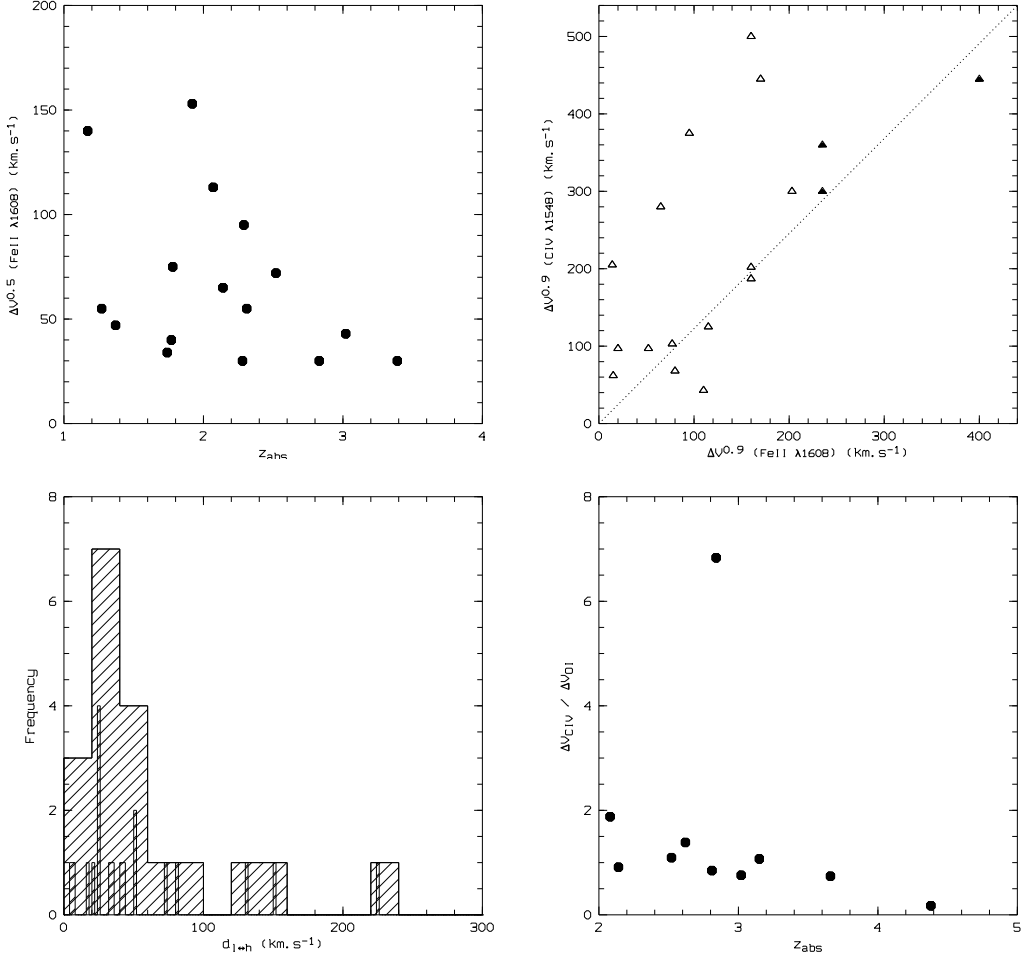

\centerline{\hbox{
\psfig{figure=0870.f14,height=7.0cm,rheight=6.5cm}
\psfig{figure=0870.f15,height=7.0cm,rheight=6.5cm}}}
\centerline{\hbox{
\psfig{figure=0870.f16,height=7.0cm,rheight=6.8cm}
\psfig{figure=0870.f17,height=7.0cm,rheight=6.8cm}}}
\caption[]{{\sl Upper panels:} velocity broadening of the \ion{Fe}{ii} lines
measured at an optical depth $\tau_{\nu}$~$\sim$~0.7 versus redshift without
considering the peculiar systems discussed in Sect. 5.2 ({\it upper left}) and
velocity broadening of the \ion{C}{iv}$\lambda$1548 line versus that of the
\ion{Fe}{ii} lines  measured at $\tau_{\nu}$~$\sim$~0.1 ({\it upper right});
filled triangles indicate the peculiar systems at $z_{\rm abs}$~=~2.46 toward
Q~0201+365, $z_{\rm abs}$~=~2.81 toward Q~0528--250, and $z_{\rm abs}$~=~3.15
toward Q~2233+131.
{\sl Lower panels:} histogram of the absolute velocity difference between the
strongest low and high-ionization components  ({\it lower left};
the smaller binning indicates individual systems) and
\ion{C}{iv} to \ion{O}{i} velocity broadening ratio versus redshift ({\it
lower right}).}
\label{patsf5z}
\end{figure*}

$\bullet$ {\it Q~2233+131 at $z$~=~3.151:} This system is peculiar for several
reasons. Djorgovski et al. (\cite{djor96}) have detected Ly$\alpha$ emission
from this system. The emission is redshifted from the main low-ionization
absorption by about 270~km~s$^{-1}$ (Lu et al. \cite{lu97}). Since the
\ion{Fe}{ii}$\lambda$1608 line does not reach $\tau_{\nu}$~=~0.7 at the
resolution of the Keck data and thus
$\Delta V_{FeII}^{0.5}$~$\sim$~0~km~s$^{-1}$, we could only derive
$\Delta V_{FeII}^{0.9}$~$\sim$~235~km~s$^{-1}$. Lu et al. (\cite{lu97}) have
argued that this system is a case for which rotation is apparent in the
absorption profile. However it is clear from the \ion{O}{i}$\lambda$1302 and
\ion{Si}{ii}$\lambda$1526 lines that this system is composed of three well
detached sub-systems spread over 300~km~s$^{-1}$ and centered at $-$270,
$-$190 and $-$80~km~s$^{-1}$ from the Ly$\alpha$ emission. There is no
edge-leading pattern as those emphasized by Prochaska \& Wolfe
(\cite{proc97b}). Instead the two sub-systems at $-$270 and $-$190~km~s$^{-1}$
are equally strong and could both contribute to the \ion{H}{i} damping wings.
It is interesting to note as well that the \ion{C}{iv} absorption does not
correlate with the low-ionization absorptions. The maximum of the \ion{C}{iv}
absorption occurs at $-$120~km~s$^{-1}$, exactly in the gap between the
low-ionization systems at $-$190 and $-$80~km~s$^{-1}$. This pattern is a
clear example of what could be expected from a disturbed object made up of
interacting sub-units. It is qualitatively not different from the pattern seen
in the $z_{\rm abs}$~=~2.8112 system toward Q~0528--250. Rotational motions
may be present in the sub-systems on velocity scales smaller than
100~km~s$^{-1}$.\\

%
From these comments, it seems that one has to be very careful when discussing
the kinematics of systems with large velocity broadenings. They often show 
the sub-systems that are expected if the objects are in the process 
of merging.
Therefore the claim that the CDM model should be rejected has to be considered
with caution. Indeed using hydro-simulations, Haehnelt et al. (\cite{haeh98})
have reached the same conclusion. Even 
though these investigations are not yet complete, the overall picture that 
seems to be emerging favors the idea that DLA systems are 
aggregates of dense
knots with complex kinematics rather than ordered disks. This does not rule
out however the  possibility that part of the kinematics could be due to
rotation. The correlation shown in Fig.~\ref{pats_f} (lower left panel) is
suggestive of rotational motions in sub-systems on scales smaller than
$\Delta V$~$<$~150~km~s$^{-1}$.

Finally, although the sample is certainly too small to draw any firm
conclusion, it is interesting to note that the mean velocity broadening of
\ion{Fe}{ii} at $\tau_{\nu}$~=~0.7 tends to decrease with redshift (see
Fig.~\ref{patsf5z}; upper left panel) {\sl if we do not include the peculiar
systems mentioned in Sect. 5.2}. If we split the sample in two sub-samples
containing the same number of systems, the mean velocity broadening is 80 and
50~km~s$^{-1}$ for $z$~$<$~2.2 and $z$~$>$~2.2 respectively. This behavior is
expected in CDM models in which disc galaxies form as gas cools and forms
stars at the centres of the dark matter haloes (e.g. Kauffmann \cite{kauf96}).
\subsection{The high-ionization phase}
The examples of absorption profiles simulated by Haehnelt et al.
(\cite{haeh98}) suggest that the comparison of the velocity profiles of high
(e.g. \ion{C}{iv}) and low (\ion{Si}{ii} or \ion{Fe}{ii}) ionization species
may give clues on the nature of the objects associated with DLA systems.
Although Fig.~6 of Haehnelt et al. (\cite{haeh98}) is drawn for illustrative
purpose, it is unclear to us whether the simulations are able or not to
reproduce observed individual cases. Contrary to what happens for \ion{Fe}{ii}
at $\tau_{\nu}$~=~0.7, there is no correlation between $\eta_{\rm
\ion{C}{iv}}$ and $\Delta V_{\rm \ion{C}{iv}}$ at $\tau_{\nu}$~=~0.1 (see
Fig.~\ref{pats_f}; lower right panel). This is expected if the high-ionization
phase has a much more disturbed kinematical field than the low-ionization
phase (see also Turnshek et al. \cite{turn89}) and indeed,
$\Delta V^{0.9}_{\rm \ion{C}{iv}}$ is most of the time larger than
$\Delta V^{0.9}_{\rm \ion{Fe}{ii}}$. However, there is a trend for
$\Delta V^{0.9}_{\rm \ion{C}{iv}}$ and $\Delta V^{0.9}_{\rm \ion{Fe}{ii}}$ both
measured at $\tau$~=~0.1 to be correlated (Fig.~\ref{patsf5z}).

The velocity differences between the maximum of the optical depth in 
\ion{C}{iv} and \ion{Fe}{ii} is most often smaller than 60~km~s$^{-1}$
(Fig.~\ref{patsf5z}). Close inspection of the spectra shows that the
kinematics of high and low-ionization species are similar. Exceptions are the
$z_{\rm abs}$~=~2.827 system toward Q~1425+606, in which the main \ion{C}{iv}
absorption corresponds to a high velocity sub-component, the
$z_{\rm abs}$~=~3.151 system toward Q~2233+131, in which the strongest
\ion{C}{iv} absorption is located in a gap between two low-ionization
sub-components, and the $z_{\rm abs}$~=~1.776 system toward Q 1331+170, in
which the \ion{C}{iv} profile is much more extended in velocity space than the
low-ionization profiles. Such behaviors can be recognized in the simulated
spectra of Haehnelt et al. (\cite{haeh98}) but should be investigated in more
details.

It is interesting to note that the
$\Delta V^{0.9}_{\ion{C}{iv}}$/$\Delta V^{0.9}_{\ion{O}{i}}$ ratio seems to
decrease with increasing redshift (see Fig.~\ref{patsf5z}). A very large value
for this ratio is found for the peculiar $z_{\rm abs}$~=~2.844 system toward
Q~1946+769 
which shows \ion{C}{iv} absorptions symmetrically placed on both sides of the
low ionization absorption. This general trend is consistent with  
kinematics to be more disturbed with decreasing redshift.

\section{Conclusions}
We have presented high resolution data for five DLA systems analyzing the
velocity profiles and the column densities.
The systems at $z_{\rm abs}$~=~3.0248 toward Q~0347--383 and $z_{\rm
abs}$~=~2.5226 toward Q~1213+093 have similar abundance patterns. Especially
the nitrogen to silicon relative abundance is smaller
than the solar value (which is true also for the system at
$z_{\rm abs}$~=~2.6184 toward Q~0913+072). 
The system at
$z_{\rm abs}$~=~2.6184 toward Q~0913+072 is the most metal-deficient DLA
system known by now. 
We find [O/H] $\approx$ $-$2.7 and [Fe/Si]~$\le$~$-$0.8. A better
limit on iron is needed. Indeed it would be difficult to reconcile this
pattern with standard chemical evolution models. It is interesting to note
that this system has a metal content similar to that of Ly$\alpha$ clouds with
log~$N$(\ion{H}{i})~$\sim$~15 (Songaila \& Cowie \cite{song96}).

From a sample of 26 DLA systems gathered from the literature, we show that 
the velocity
widths of the \ion{Si}{ii} $\lambda$1808 and \ion{Fe}{ii}$\lambda$1608 
absorptions
are closely correlated. This suggests that the physical conditions in the
neutral phase are quite homogeneous and that large relative metallicity
variations and thus large depletions into dust grains are unlikely. We discuss
the systems with \ion{Fe}{ii} velocity widths larger than 200~km~s$^{-1}$ and
conclude that they are most of the time composed of several sub-systems.
We argue, from the distribution of \ion{Fe}{ii} velocity widths and the
correlation between the asymmetry parameter and the velocity width, that
rotation motions may be present in sub-systems on scales smaller than
150~km~s$^{-1}$. The velocity width of the strongest part of the \ion{Fe}{ii}
absorption decreases with increasing redshift. This suggests that the neutral
regions get denser and exhibit faster motions with time.
%
\acknowledgements{We thank Max Pettini for sharing
information prior to publication and the referee, Giovanni Vladilo, for
insightful comments.}

%
%
\end{document}